\documentclass[12pt]{article}
\usepackage[T1]{fontenc}
\usepackage{amsfonts}
\usepackage{amsmath}
\usepackage[dvips]{graphicx,color}
\begin{document}
\title{Finite temperature regularization} 
\author{C.~D.~Fosco$^a$ 
and F.~A.~Schaposnik{\thanks{Associated with CICPBA}}$\,~^b$ \\
  {\normalsize\it $^a$Centro At{\'o}mico Bariloche - Instituto Balseiro,}\\
  {\normalsize\it Comisi{\'o}n Nacional de Energ{\'\i}a At{\'o}mica}\\
  {\normalsize\it 8400 Bariloche, Argentina.}\\
  {\normalsize\it $^b$Departamento de F{\'\i}sica, Universidad Nacional 
  de La Plata}\\ 
  {\normalsize\it C.C.~67, (1900) La Plata, Argentina}}%
\date{\hfill}
\maketitle
\begin{abstract}
\noindent We present a non-perturbative regularization  scheme for 
Quantum Field Theories which amounts to an embedding of the original
unregularized theory into a spacetime with an extra compactified
dimension of length $L \sim \Lambda^{-1}$ (with $\Lambda$ the ultraviolet cutoff),
plus a doubling in the number of fields, which satisfy different
periodicity conditions and have opposite Grassmann parity.  The
resulting regularized action may be interpreted, for the fermionic
case, as corresponding to a finite-temperature theory with a
supersymmetry, which is broken because of the boundary conditions.
We test our proposal both in a perturbative calculation (the vacuum
polarization graph for a $D$-dimensional fermionic theory) and in a
non-perturbative one (the chiral anomaly).
\end{abstract}
\section{Introduction}\label{sec:intro}
The ultraviolet (UV) regularization, is a procedure that plays a
fundamental role in the very construction of a QFT, as a crucial step
in the process of removal of UV divergences from its physical
predictions.  Besides this technical aspect, the use of a
regularization does have sometimes an important impact also on the
resulting physics. For example, if a symmetry of the model cannot be
preserved by {\em any\/} regularization method, this results in the
anomalous breaking of that symmetry, which manifests itself even at
the level of the renormalized theory~\cite{anom1}. This interplay
between classical symmetry transformations and quantum anomalies is
perhaps best understood in the context of the path integral
method~\cite{Fujikawa1,Fujikawa2}.

For the particular case of the chiral symmetry, the importance of the
anomalies, both for the global and local cases, should hardly need to
be emphasized: anomalies put constraints for model-building when the
symmetry correspond to the gauge current, and they modify the naive
predictions of current algebra for the global case~\cite{current}.

The importance of having a {\em non-perturbative\/} regularization
(like the lattice) stems from the fact that it makes it possible to
use many different approximation methods, whose application to the
usual perturbative context could be at least problematic. Of course, a
non-perturbative regularization, to be useful, should preserve as many
symmetries as possible, and they should be realized on a {\em local\/}
regularized action. In the case of theories with anomalies, the
construction of a non-perturbative regularization is a non-trivial
task since, by definition, a regularization introduces a major change
in the short distance properties of the theory, precisely the region
where theories with anomalies are most sensitive.

In this paper we present a non-perturbative regularization method
which may be formulated in terms of a higher-dimensional extension of
the model, at finite-temperature.  The original fields in the model
have to be immersed in a spacetime with an extra finite dimension, and
for the fermionic case a supersymmetry is trivially implemented by a
doubling in the number of degrees of freedom, adding to each physical
field an unphysical partner which is Grassmann but has the same action
as the fermion~\footnote{Of course, this means that the unphysical
  field violates the spin-statistics theorem, as a Pauli-Villars
  regulator does.}. This supersymmetry, exact at the level of the
Lagrangian is, however, broken because the fields have different
boundary conditions in the extra coordinate (which is bounded), whose
length is of the order of the inverse of the UV momentum cutoff.
Moreover, the non-anomalous symmetries of the theory are manifest and,
as we shall see for the regularization of fermionic determinants, the
method may be understood as an implementation of the ideas 
developed
in~\cite{Narayanan:wx}-\cite{Frolov:ck},
namely, the use of an infinite number of Pauli-Villars 
regulators~\cite{PV}-\cite{ZJ}. Our main point is to give a concrete realization
of the regularized action in terms of an action, which should then
have a compactified coordinate for the masses of the regulator fields
to emerge naturally as the corresponding discrete momenta.

For the bosonic case, on the other hand, the extension is implemented
by the addition of a bosonic partner with a negative sign in the
quadratic part of its Euclidean action, a feature also present in the
standard Pauli-Villars method, when applied to a (bosonic) scalar
field theory~\cite{ZJ}. Of course, this means that, in the canonical
quantization framework, the norm of the corresponding modes must be
negative, to avoid negative energies.  Besides, the regulator field
has antiperiodic boundary conditions in the imaginary time direction.
Also for this case, the non-perturbative regularization bears a
resemblance to the corresponding Pauli-Villars method, when the number
of regulators becomes infinite.

The use of higher dimensional representations for fermionic
determinants is a well-known idea; indeed it is the basis of the
celebrated `overlap' formalism~\cite{neuber1}, based on Kaplan's
idea~\cite{kaplan} to represent a $D$ dimensional chiral determinant
by a $D+1$ Dirac theory with a non-homogeneous mass. However, our
motivation here is to find a representation of the {\em regularized\/}
fermionic determinant in the continuum.

The existence of a compact length for the extra dimension makes it
possible to use a Matsubara-like formalism, as in finite-temperature
QFT. This allows for the derivation of some results that can be simply
adapted from one context to the other. For example, the finiteness of
the regularization procedure may be seen as a result of the well-known
fact that in finite-temperature QFT the UV divergences are the same as
in QFT at $T=0$, in combination with the property that the
zero-temperature limit of the regularized theory is trivial: for the
fermionic case, the supersymmetry is exact at $T=0$, while for the
bosonic case there is no effective propagation in that limit.

The organization of this paper is as follows: in
section~\ref{sec:fdet} we motivate and define the regularization by a
study of the particular case of a fermionic determinant in an external
field. In section~\ref{sec:bos}, we extend the method to a bosonic
QFT: the self-interacting real scalar field. In
section~\ref{sec:pert}, we perform perturbation theory tests on the
method, in order to check and understand some of the properties of the
regularization in concrete examples.  Some non-perturbative tests,
like the issue of chiral anomalies are dealt with in
section~\ref{sec:anom}, and section~\ref{sec:conc} contains our conclusions.
\section{Fermionic theory}\label{sec:fdet}
We begin with the particular case of a fermionic determinant in an
external gauge field, which, as we shall see, serves both as
motivation and test for the method.  To this end, we first introduce
the partition function ${\mathcal Z}_f(A)$ for the unregularized
theory, which is defined by:
\begin{equation}\label{eq:defzf}
{\mathcal Z}_f(A)\;=\; \int {\mathcal D}\psi {\mathcal D}{\bar\psi} \;
\exp \left[- S_f({\bar\psi},\psi;A)\right] 
\end{equation}
where $S_f$ denotes the unregularized Euclidean action for a massless
Dirac field in $D$ dimensions, i.e.,
\begin{equation}\label{eq:defsf}
S_f({\bar\psi},\psi;A) \;=\;\int d^Dx \; {\bar\psi}{\mathcal D}_c \psi 
\end{equation}  
where ${\mathcal D}_c$ is the massless Dirac operator:
\begin{equation}\label{eq:defdc}
{\mathcal D}_c \;=\; \not \! \partial + \not \!\! A
\end{equation}
and the $\gamma$ matrices satisfy the conventions:
\begin{equation}\label{eq:gconv}
\gamma_\mu^\dagger \,= \, \gamma_\mu \;\;,\;\;\;
\{ \gamma_\mu , \gamma_\nu \} \;=\; 2 \, \delta_{\mu\nu}\;.
\end{equation}
The gauge connection $A$ verifies
\begin{equation}
A_\mu \;=\;-A_\mu^\dagger \;\equiv\;\left\{
\begin{array}{c}
i {\mathcal A}_\mu \;\;{\rm in \,the\, Abelian\, case}\\
{\mathcal A}_\mu^a \tau_a \;\; {\rm in\, the\, non\, Abelian\, case}
\end{array}
\right.\;,
\end{equation}
where $\tau_a$ are (anti-hermitian) generators of the Lie algebra of
the non-Abelian gauge group, and both ${\cal A}_\mu$ and ${\cal
  A}_\mu^a$ are real.  This implies the anti-hermiticity of ${\mathcal
  D}_c$ in Euclidean space, a property that will be taken into account
for the construction of the regularized theory.

To introduce the non-perturbative regularization, we define a
`regularized' Dirac operator ${\mathcal D}$, a function of the
unregularized operator ${\mathcal D}_c$, through the equation
\begin{equation}\label{eq:defd}
\frac{i \mathcal D}{M}\;=\; f(\frac{i{\mathcal D}_c}{M})
\end{equation}
where $M$ is a constant with the dimensions of a mass, playing the
role of an UV cutoff and the function $f$ has to be chosen in order to
tame the UV behaviour of the Dirac operator. Since the UV properties
manifest themselves in the large eigenvalues of ${\mathcal D}_c$, when
considered as a function of a real variable $x$, \mbox{$f:{\mathbb R}
  \to {\mathbb R}$} should verify
\begin{equation}\label{eq:fprop}
x \to 0 \;\;\Rightarrow \;\; f(x) \,\sim\, x \;\;,\;\;
x \to \pm \infty \;\;\Rightarrow \;\; f(x) \,\to \, \pm 1\;.
\end{equation}
Note that the first relation amounts to requiring the regularized operator
to behave like the unregularized one when the eigenvalues are small, while 
the second condition implies that ${\mathcal D}$ is bounded. Indeed,
the spectrum of ${\mathcal D}$ is confined to the range $[-M,M]$. Besides, 
the function $f$ should be one to one, in order to preserve some important 
properties of the spectrum, as the degeneracy of each eigenvalue. 
A convenient choice for $f$ satisfying all of these constraints 
is: 
\begin{equation}\label{eq:deff}
\frac{i \mathcal D}{M}\;=\; \tanh (\frac{i{\mathcal D}_c}{M})\;,
\end{equation}
which, of course, would yield a non-local theory if ${\mathcal D}$
were used to build a $D$-dimensional theory. Rather than working with
this non-local expression, we shall instead consider an equivalent
formulation of this regularized theory where the non-locality is
traded 
\underline{
for the existence
  of a compactified extra}\\ 
   \underline{
dimension, plus an extra unphysical field}. 
 A first step in that
direction is to use an infinite product representation for the $\tanh$
function,
\begin{equation}\label{eq:infprd}
\tanh(x) \;=\; \xi \, x \,  \prod_{n=1}^{\infty} 
\frac{x^2 + n^2 \pi^2}{x^2 + ( n-\frac{1}{2})^2 \pi^2} 
\end{equation}
where $\xi = \prod_{n=1}^\infty (1 - \frac{1}{2 n})^2$.  By the use of some
straightforward algebra, we may insert (\ref{eq:infprd}) in
(\ref{eq:deff}) to write the regularized Dirac operator as:
\begin{equation}\label{eq:infprd1}
{\mathcal D} \;=\; \,\xi \, \prod_{n=-\infty}^{+\infty}
\left[ \frac{{\mathcal D}_c + \pi n M}{{\mathcal D}_c +
(n+\frac{1}{2}) \pi M}\right]\;, 
\end{equation}
a form which already suggests the use of a higher dimensional
representation and the introduction of fields of opposite statistics
to obtain a local version of the determinant of ${\mathcal D}$.
Indeed, we note that the regularized partition function ${\mathcal
  Z}_f^{reg}$, defined in terms of ${\mathcal D}$, may be written as
follows:
$$
{\mathcal Z}_f^{reg}= \int {\mathcal D}\psi {\mathcal D}{\bar\psi} \,
\exp \left[-\int d^Dx  \; {\bar\psi} {\mathcal D}\psi  \right] 
$$
\begin{equation}\label{eq:infprd2}
=\; \exp \sum_{-\infty}^{+\infty} \left\{ {\rm Tr} \ln [{\mathcal
D}_c + \pi n M] - {\rm Tr} \ln [{\mathcal D}_c + (n+\frac{1}{2}) \pi M]
\right\}\;.
\end{equation}
It should be evident that the sum over $n$ of the $D$-dimensional
traces may also be interpreted as $D+1$-dimensional traces for a
Finite Temperature theory, in the Matsubara formalism, and with
translation invariance along the imaginary time coordinate. In this
finite-temperature language, we have $\beta=\frac{2}{M}$ (or $T =
\frac{M}{2}$) and the two operator traces are associated with two
fields with opposite (odd/even) Grassmann character which have to be
integrated with opposite (periodic/antiperiodic) boundary conditions.
Indeed, the two fields can be naturally associated to the Matsubara
frequencies:
\begin{equation}\label{eq:defmsb}
\omega^{(+)}_n \;=\; \frac{2 \pi}{\beta}  n  \;=\; \pi n M \;\;,\;\;\;
\omega^{(-)}_n \;=\; \frac{2 \pi}{\beta}  (n+\frac{1}{2}) \;=\; \pi
(n+\frac{1}{2}) M \;,
\end{equation}
where $\omega^{(+)}_n$ and $\omega^{(-)}_n$ correspond to periodic and
antiperiodic boundary conditions, respectively.  According to this
interpretation, we may then write the partition function as a
functional integral over two Dirac fields in $D+1$ dimensions,
$\Psi^{(+)}(\tau,x)$ and $\Psi^{(-)}(\tau,x)$, associated to the two 
different types of boundary conditions (bc) and with 
opposite Grassmann character,
\begin{eqnarray}
 \Psi^{(+)}(\tau,x) &:& {\rm periodic ~bc, ~Grassmann~odd} \nonumber\\
 \Psi^{(-)}(\tau,x) &:& {\rm antiperiodic ~bc, ~Grassmann~even}
 \nonumber
 \end{eqnarray}
The resulting partition function reads
\begin{equation}\label{eq:regfint}
{\mathcal Z}_f^{reg} = \int {\mathcal D} \Psi^{(+)} {\mathcal D} \Psi^{(-)}
\, {\mathcal D} {\bar\Psi}^{(+)} {\mathcal D} {\bar\Psi}^{(-)}
\, \exp\left[ - {\mathcal S}_f^{reg} ({\bar\Psi}^{(+)},{\bar\Psi}^{(-)},\Psi^{(+)},
\Psi^{(-)}; A) \right] 
\end{equation} 
with
\begin{equation}\label{eq:defsfreg}
{\mathcal S}_f^{reg}=\int_{-\frac{1}{M}}^{+\frac{1}{M}} d \tau \int
d^Dx \, \left[ {\bar\Psi}^{(+)}(- i \partial_\tau + {\mathcal D}_c) \Psi^{(+)}
+{\bar\Psi}^{(-)}(- i \partial_\tau + {\mathcal D}_c ) \Psi^{(-)} \right] 
\end{equation}
Notice that the integration rules associated with $\Psi^{(+)}$ coincide with
those  to be imposed on ghosts in finite temperature gauge theories. In this
sense, one could think on conditions imposed to
 $\Psi^{(-)}$ as those corresponding to ghosts of ghosts.

Concerning action (\ref{eq:defsfreg}), it should be noted that it is
not covariant when considered as a $D+1$ dimensional object.  This is
not a problem, of course, since real physics corresponds to $D$
dimensions, where the theory is indeed invariant. Besides, the fact
that there is a finite range for the imaginary time $\tau$ coordinate
already breaks (explicitly) the symmetry between $\tau$ and the remaining
$D$ coordinates, denoted collectively by $x$. This breaking is a usual
phenomenon in Finite Temperature QFT, and may be traced to the fact
that there is a preferential reference system, namely, the `thermal
bath'.  However, had the original theory been defined on an even
number of dimensions, i.e., $D = 2 n$, the corresponding hermitian
chirality matrix $\gamma_s$ ($\gamma_s = \gamma_5$ when $D=4$) could have been used
in order to obtain an equivalent $D+1$-invariant looking expression
for the Lagrangian.  Indeed, it is sufficient to note that, both for
$\Psi^{(+)}$ or $\Psi^{(-)}$, we have
$$
{\bar \Psi} (- i \partial_\tau + {\mathcal D}_c ) \Psi \;=\; {\bar \Psi} (- i \gamma_s) i
\gamma_ s (- i \partial_\tau + \not \!\! D) \Psi
$$
\begin{equation}\label{eq:covers}
\;\equiv\; {\tilde \Psi} \Gamma_\alpha D_\alpha  \Psi \;.
\end{equation} 
where, in the last line, the index $\alpha$ runs from $0$ to $D$, and we
shall alternatively use the notation $\tau$ or $x^D$ for the extra
coordinate, depending on the context.  We have introduced the Dirac
matrices $\Gamma_\alpha$ in $D+1$ dimensions in such a way that $\Gamma_\mu = i \gamma_s
\gamma_\mu$, $\forall \mu$ such that $0 \leq \mu \leq D-1$, and $\Gamma_D \equiv \Gamma_\tau \equiv \gamma_s$.

Of course, $\Gamma_\alpha^\dagger = \Gamma_\alpha$, and
\begin{equation}
\{\Gamma_\alpha \,,\, \Gamma_\beta \} \;=\; 2 \, \delta_{\alpha\beta}
\;\;\;,\;\;\; \forall \alpha, \beta = 0,1,\ldots, D\;.
\end{equation}
Also, the $\alpha=D$ component of $A_\alpha$ vanishes.  In the last line of
(\ref{eq:covers}) we have used the notation ${\tilde \Psi} \equiv {\bar\Psi} i
\gamma_s = \Psi^\dagger \Gamma_0$, i.e., ${\tilde \Psi}$ is the natural definition for
${\bar\Psi}$ in the $D+1$ covariant representation. With this in mind, we
shall often write also ${\bar\Psi}$ (rather than ${\tilde\Psi}$) when
working in the $D+1$ covariant representation, since the meaning of
the bar should be clear from the context.

The changes to introduce when the original theory is massive are quite
straightforward, since they stem from the replacement ${\mathcal D}_c
\to{\mathcal D}_c + m $, where $m$ is the fermion mass, in
(\ref{eq:defd}). However, when the $D+1$ covariant notation is used,
we note that the physical mass $m$ will arise as a constant gauge
potential $A_\tau = i m$ in the extra coordinate, namely,
\begin{equation}\label{eq:massivesfreg}
{\mathcal S}_f^{reg}(m) =\int d^{D+1}x \, \left[ {\bar\Psi}^{(+)} \Gamma_\alpha D_\alpha(m) \Psi^{(+)}
+ {\bar\Psi}^{(-)}\Gamma_\alpha D_\alpha(m) \Psi^{(-)} \right] 
\end{equation}
where $d^{D+1}x \equiv d\tau d^Dx$, and $D_\tau \equiv \partial_\tau + i m$.  The fact that the
$\tau$ coordinate is compact implies that the constant $m$ cannot be
gauged away, unless a twist is introduced for the fermions. This
twisting would carry, of course, the same physical content as the
constant gauge field.

The regulated Lagrangian in $D+1$ dimensions is supersymmetric, in the
sense that it is invariant under the global transformations:
\begin{eqnarray}\label{eq:glsusy}
\delta\Psi^{(+)} &=& i \xi  \Psi^{(-)} \;\;\;\;\;  \delta{\bar \Psi}^{(+)} \;=\; - i {\bar
  \Psi}^{(-)} \xi \nonumber\\
\delta\Psi^{(-)} &=& i \xi  \Psi^{(+)} \;\;\;\;\;  \delta{\bar \Psi}^{(-)} \;=\; - i {\bar
  \Psi}^{(+)} \xi \;,
\end{eqnarray}
where $\xi$ is a real Grassmann variable. One should 
expect  that this symmetry is 
broken because of the different boundary conditions for the $(+)$ and $(-)$ fields. 

The symmetry (\ref{eq:glsusy}) may be presented in a more explicit way by the 
introduction of a superfield notation:  we introduce a pair of conjugate Grassmann 
variables $\theta$, ${\bar \theta}$, and use the notation $\chi$ for a fermionic 
superfield defined as follows:
\begin{eqnarray}
\chi (\tau,x,\theta) &=& \Psi^{(+)} \,+\, \theta \Psi^{(-)} \nonumber\\ 
{\bar \chi} (\tau,x,\theta) &=& {\bar \Psi}^{(+)} \,+\,{\bar \Psi}^{(-)} {\bar \theta}\;, 
\end{eqnarray} 
namely, $\chi$ is `analytic' in the Grassmanian sense: $\partial_{\bar \theta} 
\chi = 0$.  The regularized action may then be written in terms of
this analytic field in the following way:
\begin{equation}\label{eq:super}
{\mathcal S}_f^{reg}=\int_{-\frac{1}{M}}^{+\frac{1}{M}} d \tau \int
d^Dx \int d\theta d{\bar \theta} \, e^{-\theta {\bar\theta}} \; {\bar\chi} \; \Gamma_\alpha D_\alpha \; \chi 
\end{equation}
where we note the presence of the exponential factor $e^{-\theta {\bar\theta}}$,
as usual for the inner product between analytic functions of a
Grassmann variable. The boundary conditions on the compact coordinate
may now be written in terms of $\chi$:
\begin{equation}
\chi(\beta,x,\theta) \;=\; \chi(0,x,-\theta) \;\;\;\;,\; \forall x.
\end{equation}

We conclude this section with an alternative (also finite-temperature)
representation for the regularized fermionic determinant. It should be
evident that the logarithm of the regularized
partition function (\ref{eq:regfint}), may also be written in terms of 
traces over anti-periodic fields, namely
$$
\ln {\mathcal Z}_f^{reg} \;=\; {\rm Tr}^{(-)}\ln \left[ 
\Gamma_D(\partial_\tau + i \frac{\pi}{\beta}) + \Gamma_\mu D_\mu
\right]
$$
\begin{equation}\label{eq:alt1}
-\, {\rm Tr}^{(-)}\ln \left[ \Gamma_D\partial_\tau + \Gamma_\mu D_\mu \right] 
\end{equation}
where the $(-)$ over the traces indicates that for both fields it is
antiperiodic, while the periodicity for the first term has been traded
for the presence of a constant gauge field. This may also be written
in the equivalent way: 
\begin{equation}\label{eq:alt2}
\ln {\mathcal Z}_f^{reg} \;=\; \int_0^{\frac{\pi}{\beta}} d a
\frac{\partial}{\partial a} \, {\rm Tr}^{(-)}\ln \left[ \Gamma_\alpha D_\alpha (a)\right] 
\end{equation}
where 
\begin{equation}
D_\mu(a) = D_\mu \;\;\;,\;\;\;D_D (a) = \partial_\tau + i a \;.
\end{equation}
By taking the derivative with respect to $a$, in (\ref{eq:alt2}), we
see that the equation becomes an integral over $a$ of 
the thermal average of the conserved charge $Q(a)$, corresponding to the $D+1$
dimensional current for a (single) Dirac fermion. Namely,
$$
\ln {\mathcal Z}_f^{reg} \;=\; \int_0^{\frac{\pi}{\beta}} d a  {\rm
  Tr}^{(-)}\left[ i \Gamma_\tau ( \Gamma_\alpha D_\alpha (a))^{-1}\right] 
$$
\begin{equation}\label{eq:alt3}
\;=\; \int_0^{\frac{\pi}{\beta}} d a \langle Q(a) \rangle 
\end{equation}
where the charge $Q(a)$ depends of course on the value of the constant
$a$ and on the external field $A_\mu(x)$. When $D$ is even, it is
interesting to see that $\langle Q(a) \rangle$ may be decomposed into its parity
breaking and parity conserving parts, yielding the same decomposition
for the $D$-dimensional regularized effective action.

When $M \to \infty$ ($\beta \to 0$), we see that $\langle Q(a) \rangle$ becomes just the charge
corresponding to the Chern-Simons current in $D+1$ dimensions, thus
reproducing the well-known relation between gauge topological terms in
even and odd dimensions.  It is amusing to see that this
regularization somehow preserves a relation like that, even when the
theory is regulated.

When the original theory is massive, the only change is in the
integration range for the $a$ integral in (\ref{eq:alt3}), i.e., 
\begin{equation}\label{eq:alt4}
\ln {\mathcal Z}_f^{reg}(m) \;=\; \int_{m}^{\frac{\pi}{\beta}+m} d a \langle Q(a) \rangle \;. 
\end{equation}

\section{Bosonic theory}\label{sec:bos}
Let us adapt here the previously introduced idea to the case of a
bosonic action. We take as the essential property of the method 
to preserve by such a generalization, not the functional
relation between two Dirac operators, but rather the finite
temperature interpretation. Indeed, the latter is a framework that can
be implemented for any model, regardless of the spin and internal
symmetry group of the fields, and its precise realization may be
inferred from the Pauli-Villars method, by introducing the proper
generalization.

To be more precise, we consider a real scalar field $\varphi$ in $D$
Euclidean dimensions, described by the action:
\begin{equation}\label{eq:defsscal}
S_D\;=\;\int d^Dx \; \left[ \frac{1}{2} \partial_\mu \varphi \partial_\mu \varphi \,+\,
\frac{1}{2} m^2 \varphi^2  \,+\, V(\varphi)\right]\;. 
\end{equation}
Guided by the fermionic case, we introduce now two scalar fields
$\Phi^{(\pm)}(\tau,x)$ and a finite temperature action (with $\beta =
\frac{2}{M}$) in $D+1$ dimensions, with periodic boundary conditions
for the field $\Phi^{(+)}(\tau,x)$:
\begin{equation}\label{eq:fexp}
\Phi^{(+)}(\tau,x)\;=\; \beta^{-\frac{1}{2}}\,
\sum_{n=-\infty}^{+\infty} \Phi^{(+)}_n (x) \, e^{i \omega^{(+)}_n \tau}
\end{equation}
where $\omega^{(+)}_n = \frac{2\pi n}{\beta}$ and 
\begin{equation}
\Phi^{(+)}_n (x)\;=\; \beta^{-\frac{1}{2}} \, \int_{-\beta/2}^{+\beta/2}
d\tau \, \Phi^{(+)}(\tau,x) \, e^{- i \omega^{(+)}_n \tau} \;.
\end{equation}
and antiperiodic boundary conditions for $\Phi^{(-)}$.

In particular, the zero mode of $\Phi^{(+)}(\tau,x)$ is assumed to be
proportional to $\varphi(x)$, i.e.,
\begin{equation}
\Phi^{(+)}_0 (x)= \beta^{-\frac{1}{2}} \varphi(x)\;.
\end{equation}
On the other hand, inspired by the Pauli-Villars method in its scalar
field version~\cite{ZJ}, we may introduce the regularized action ${\mathcal
  S}_{D+1}$ as follows:
$$
{\mathcal S}_{D+1}[\Phi^{(+)},\Phi^{(-)}] \;=\;
\int_{-\frac{1}{M}}^{+\frac{1}{M}} d \tau \!\!\int \! d^Dx \, \left\{ 
\frac{1}{2} [ \partial_\alpha\Phi^{(+)}  \partial_\alpha\Phi^{(+)} + m^2 \Phi^{(+)}\Phi^{(+)}] 
\right.
$$
\begin{equation}\label{eq:defsgen}
\left. - \frac{1}{2} [ \partial_\alpha\Phi^{(-)}  \partial_\alpha\Phi^{(-)} + m^2 \Phi^{(-)}\Phi^{(-)}]
+ V( \Phi^{(+)} + \Phi^{(-)})
\right\} \;.
\end{equation}

Since there is no free propagator connecting $\Phi^{(+)}$ to $\Phi^{(-)}$, an 
application of the Wick theorem to a given Green's function yields the
result that any diagram in the perturbative expansion can be
built in the following way: replace in every diagram of the unregularized 
theory the free propagator by the sum of the propagator for $\Phi^{(+)}$ and the
propagator for $\Phi^{(-)}$ (which differ in a global sign and in
their boundary conditions). Thus the `regularized propagator' $G^{reg}$ is
\begin{equation}
D^{reg} \;=\; \langle  \Phi^{(+)}   \Phi^{(+)} \rangle + \langle  \Phi^{(-)}  \Phi^{(-)} \rangle \;.
\end{equation}
Taking into account the different boundary conditions and the minus
sign in front of the $(-)$ action, we see that in momentum space the 
sum of the propagators becomes, at large momenta, exponentially damped. Namely,
\begin{equation}
  D^{reg}(k) \;\sim \; exp[-\beta |k|] \;.
\end{equation}

This explains the finiteness of the theory; we also need to be sure
that the unregularized theory is recovered when $M \to \infty$. This property
is indeed also true, since in this limit the antiperiodic fields
$\Phi^{(-)}$ become infinitely massive, as well as all the non-zero modes
of $\Phi^{(+)}$. Thus the theory is dimensionally reduced to the zero
Matsubara frequency mode, which, by construction, is tantamount to
$\varphi(x)$, and of course has the original action $S_D$.

\section{Some perturbative tests}\label{sec:pert}
In this section we shall apply the regularization method to the
derivation of some perturbative results. We shall begin by considering
the $D$-dimensional vacuum polarization graph for the fermionic
determinant, to one-loop order, and for the Abelian gauge field case.

The regularized vacuum polarization function ${\tilde \Pi}_{\mu\nu}$ can, in
this regularization, be written as the difference between the
contributions corresponding to periodic and antiperiodic boundary
conditions.  Besides, the gauge field is independent of the $\tau$
coordinate, thus we may write
\begin{equation}\label{eq:pi}
{\tilde \Pi}_{\mu\nu}(k)\;=\; \int \frac{d^Dp}{(2\pi)^D} \, \left[
t^{(+)}_{\mu\nu}(p,k) \,-\, t^{(-)}_{\mu\nu}(p,k) \right] 
\end{equation}
where
\begin{equation}
t^{(\pm)}_{\mu\nu}(p,k) \;=\; \sum_{n=-\infty}^{+\infty} \,
{\rm tr} \left[
\frac{1}{i \not \! p + \omega_n^{(\pm)}} \gamma_\mu 
\frac{1}{i (\not \! p + \not \! k) + \omega_n^{(\pm)}} \gamma_\nu 
\right] 
\end{equation}
and we have used the non-covariant expression for the Dirac matrices
(the final result for ${\tilde \Pi}_{\mu\nu}$ is of course independent of
the representation employed). Of course, since there is no physical
gauge field component corresponding to $\alpha=D$, we are not in principle
interested in evaluating the components of ${\tilde \Pi}_{\alpha\beta}$ involving
that index. A similar remark holds true for the independence of the
external field on $\tau$, what allows us to set all the external
Matsubara frequencies equal to zero.

Evaluating the Dirac traces for  $t_{\mu\nu}^{(\pm)}$, we see that 
\begin{equation}
t^{(\pm)}_{\mu\nu}(p,k) = {\rm tr}(I) \sum_{n=-\infty}^{+\infty} \,
\frac{-p_\mu (p+k)_\nu - p_\nu (p+k)_\mu + [p \cdot (p + k)
+ \omega^{(\pm)}_n] \delta_{\mu\nu}}{(p^2
+(\omega^{(\pm)}_n)^2)[(p+k)^2 +(\omega^{(\pm)}_n)^2 ]} 
\end{equation}
where ${\rm tr}(I)$ denotes the trace over the identity matrix in
Dirac space, i.e., the dimension of the Clifford algebra
representation.

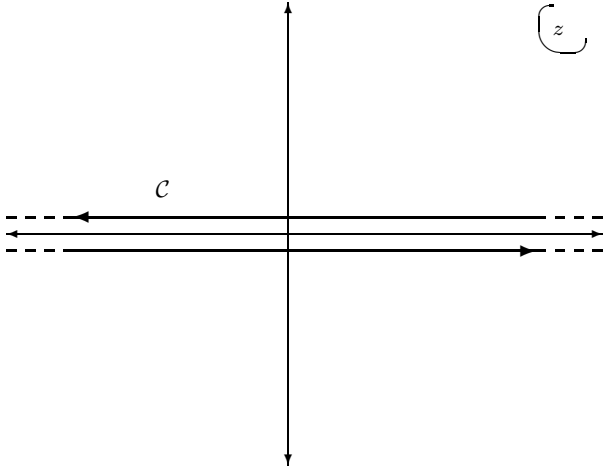
\begin{figure}[h]
\begin{center}
\setlength{\unitlength}{1823sp}%
\begingroup\makeatletter\ifx\SetFigFont\undefined%
\gdef\SetFigFont#1#2#3#4#5{%
  \reset@font\fontsize{#1}{#2pt}%
  \fontfamily{#3}\fontseries{#4}\fontshape{#5}%
  \selectfont}%
\fi\endgroup%
\begin{picture}(8166,6324)(1543,-6823)
\thinlines
{\color[rgb]{0,0,0}\put(9001,-858){\oval(378,604)[tl]}
\put(9149,-858){\oval(674,674)[bl]}
\put(9149,-1006){\oval(604,378)[br]}
}%
\thicklines
{\color[rgb]{0,0,0}\multiput(1576,-3886)(257.14286,0.00000){4}{\line( 1, 0){128.571}}
}%
{\color[rgb]{0,0,0}\multiput(8776,-3886)(257.14286,0.00000){4}{\line( 1, 0){128.571}}
}%
{\color[rgb]{0,0,0}\multiput(1576,-3436)(257.14286,0.00000){4}{\line( 1, 0){128.571}}
}%
{\color[rgb]{0,0,0}\multiput(8776,-3436)(257.14286,0.00000){4}{\line( 1, 0){128.571}}
}%
{\color[rgb]{0,0,0}\put(2476,-3886){\vector( 1, 0){6300}}
}%
{\color[rgb]{0,0,0}\put(8776,-3436){\vector(-1, 0){6300}}
}%
\thinlines
{\color[rgb]{0,0,0}\put(5401,-511){\vector( 0, 1){  0}}
\put(5401,-511){\vector( 0,-1){6300}}
}%
{\color[rgb]{0,0,0}\put(1576,-3661){\vector(-1, 0){  0}}
\put(1576,-3661){\vector( 1, 0){8100}}
}%
\put(9001,-961){\makebox(0,0)[lb]{\smash{\SetFigFont{8}{9.6}{\rmdefault}{\mddefault}{\updefault}{\color[rgb]{0,0,0}$z$}%
}}}
\put(3601,-3166){\makebox(0,0)[lb]{\smash{\SetFigFont{9}{10.8}{\rmdefault}{\mddefault}{\updefault}{\color[rgb]{0,0,0}${\mathcal C}$}%
}}}
\end{picture}
\caption{The curve ${\mathcal C}$ used in the integral
(\ref{eq:fqsum1}).}
\label{fig:curve}
\end{center}
\end{figure}

It should be remembered, before integrating out the momenta $p$, that
the regularization works only if both the $(+)$ and $(-)$
contributions are taken into account inside the integrand, i.e., the
integral cannot be distributed (as in the Pauli-Villars method). This
requires the knowledge of sums over frequencies which are not the
standard ones of the Matsubara formalism, but rather objects defined
as follows:
\begin{equation}\label{eq:fqsum}
\sigma (f) \;\equiv\; \sum_{n=-\infty}^{+\infty} \left[ f(n) -
f(n+\frac{1}{2}) \right] \;.
\end{equation}
By the same kind of trick applied in the Matsubara formalism, we may express 
the series (\ref{eq:fqsum}) as a complex contour integral,
\begin{equation}\label{eq:fqsum1}
\sigma (f) \;=\; \frac{1}{i} \oint_{\mathcal C} dz \, \frac{f(z)}{\sin
(2 \pi z)} 
\end{equation}
where ${\mathcal C}$ is the curve shown in figure~\ref{fig:curve}.

Performing the sums over frequencies according to this rule, we end up
with an expression for ${\tilde \Pi}_{\mu\nu}(k)$, which may be written
as follows:
$$
{\tilde \Pi}_{\mu\nu}(k)\;=\; \beta \, {\rm tr}(I) \, \int
\frac{d^Dp}{(2 \pi)^D}\, \frac{1}{(p+k)^2 - p^2} 
$$
$$
\,\left\{
[\frac{1}{|p| \sinh (\beta |p|)} - \frac{1}{|p + k| \sinh (\beta
|p+k|)} ]
\right.
$$
$$
[-p_\mu (p+k)_\nu - p_\nu (p+k)_\mu + p \cdot (p + k) \delta_{\mu\nu}]
$$
\begin{equation}\label{eq:pi1}
\left. \,-\,  [\frac{|p|}{\sinh (\beta |p|)} - \frac{|p+k|}{\sinh (\beta
|p+k|)} ] \delta_{\mu\nu} \right\}\;,
\end{equation}
where the UV convergence of the integral over $p$ is evident (the
symbol $|p|$ denotes the $D$-dimensional Euclidean norm).

The finite-temperature regularization is also explicitly gauge
invariant for gauge transformations in the $D$-dimensional space.
This implies the ($D$-dimensional) transversality of the
${\tilde \Pi}_{\mu\nu}$ tensor, i.e., 
\begin{equation}
k_\mu \, {\tilde \Pi}_{\mu\nu}(k) \;=\; 0\;,
\end{equation}
and we may then write ${\tilde \Pi}_{\mu\nu}(k)$ in terms of a scalar
function ${\tilde \pi}(k)$ and  a transverse projector, in the standard
fashion:
\begin{equation}\label{eq:transv}
{\tilde \Pi}_{\mu\nu}(k)\;=\; {\tilde \Pi}(k) \, (\delta_{\mu\nu}
- \frac{k_\mu k_\nu}{k^2})\;.
\end{equation} 

Taking into account (\ref{eq:transv}) and (\ref{eq:pi1}), and
performing shifts in the integration variables~\footnote{This is
  licit, since the exponential factors from the $\sinh$ render all the
  terms convergent.}, we note that
\begin{equation}
{\tilde \Pi}(k)\;=\; \frac{2 \beta {\rm tr}(I)}{D-1} \, 
\int \frac{d^Dp}{(2 \pi)^D} \, \frac{ (D-2) (|p| + k \cdot {\hat p}) -
D p}{(k^2 + 2 k \cdot p) \sinh( \beta |p|)} \;,
\end{equation}
which is a finite scalar function of $k$ for any value of $D$. The UV
behaviour of the integrand is of course the same as
in~\cite{Frolov:ck}.  It should be clear from our starting point
(\ref{eq:deff}), that convergence will be achieved for all orders,
and not just for the diagram quadratic in the external field.

\section{Chiral anomaly}\label{sec:anom}
In this section we shall restrict ourselves to the case of a massless
fermion in an even number of dimensions, in order to  observe
the emergence of the chiral anomalies in this context.
Since the finite temperature regularization introduces an extra dimension
into the game, it is not at all obvious whether the original
chiral symmetry is still meaningful or not. It is however, easy to see
that there is, indeed, a symmetry transformation which corresponds to
the chiral transformations when the regulator is removed. They are
however non-local on a scale of the order of $M$, and may be written
as follows:
$$
 \delta \Psi^{(\pm)} (\tau, x) \;=\; i \,\alpha \, \Gamma_D \, (\Gamma_\mu D_\mu
- \Gamma_D \partial_D)^{-1} \, \Gamma_\nu D_\nu \, \Psi^{(\pm)} (\tau,x)    
$$
\begin{equation}\label{eq:trans}
 \delta {\bar \Psi}^{(\pm)} (\tau, x) = i \alpha {\bar \Psi}^{(\pm)} (\tau, x) 
\Gamma_\nu D_\nu (\Gamma_\mu D_\mu - \Gamma_D \partial_D)^{-1} \, 
\Gamma_D \,. 
\end{equation}

When $M \to \infty$, the fields are dimensionally reduced and
$\tau$-independent. The transformations reduce of course to the standard
ones, since the $\partial_D = \partial_\tau$ operator yields zero when acting on a
dimensionally reduced field. When $M$ is finite, they have of course a
more complicated-looking expression, but nevertheless they leave the
regularized action invariant.

There is an effect on the integration measure: we have a
super-Jacobian, which is non-vanishing because of the different
boundary conditions (otherwise, the $(+)$ and $(-)$ contributions
would cancel).  Indeed, under the transformations (\ref{eq:trans}),
the functional integration measure
\begin{equation}
{\mathcal D}\mu \;\equiv\; {\mathcal D}\Psi^{(+)}{\mathcal D}\Psi^{(-)}
\; {\mathcal D}{\bar\Psi}^{(+)} {\mathcal D}{\bar\Psi}^{(-)}   
\end{equation}
transforms as follows:
\begin{equation}
{\mathcal D}\mu \;\to \;{\mathcal D}\mu \; {\mathcal J} 
\end{equation}
where
$$
\ln {\mathcal J} \;=\; - i \alpha \left\{ {\rm Tr} [\Gamma_D 
(\Gamma_\mu D_\mu - \Gamma_D \partial_D)^{-1} \, \Gamma_\nu D_\nu
]^{(+)}\right.
$$
\begin{equation}
\left. +\, {\rm Tr} [\Gamma_\nu D_\nu (\Gamma_\mu D_\mu - \Gamma_D \partial_D)^{-1} \, 
\Gamma_D ]^{(-)} \;-\; (+) \to (-) \right\}\;,
\end{equation}
where the $(\pm)$ indicates whether the corresponding trace has to be taken
on the space of symmetric or antisymmetric functions. A simple algebra
shows that:
$$
\ln {\mathcal J} \;=\; - i \alpha \left\{ {\rm Tr} [\Gamma_D  
(\Gamma_\nu D_\nu)^2 (\Gamma_\mu D_\mu - \Gamma_D \partial_D)^{-1}]^{(+)}
\right.$$
\begin{equation}
\left. -\; (+) \to (-) \right\}\;,
\end{equation}
and, since the gauge fields are independent of $\tau$, we may evaluate
the trace over the $\tau$ coordinate. This sum over frequencies
yields:
\begin{equation}
\ln {\mathcal J} \;=\; - 2 i \alpha \, {\rm Tr}\left[ 
\Gamma_D \frac{ \beta \Gamma_\mu D_\mu }{\sinh (\beta \Gamma_\mu D_\mu
)}\right] \;.
\end{equation}
This expression may be conveniently written in a $D$ dimensional
notation, as follows:
\begin{equation}
\ln {\mathcal J} \;=\; - 2 i \alpha \, {\rm Tr}\left[ 
\gamma_5 \, \varphi (\frac{\not \!\! D}{\Lambda}) \right]
\label{formula}
\end{equation}
where:
\begin{equation}
\varphi(x) \,=\, \frac{x}{\sinh(x)}
\end{equation}
and $\Lambda \equiv {M}/{2}$. 

Now, the function $\varphi(x)$ verifies:
\begin{equation}
\varphi (0) = 1\;\;,\;\;\;\lim_{x\to\infty} \varphi(x) = \lim_{x\to\infty} 
\varphi'(x) = \ldots = \lim_{x\to\infty} \varphi^{(k)}(x) = \ldots = 0
\;,
\end{equation}
the conditions to be imposed to regulating functions $\varphi(x)$ in order
to give a $\varphi$-independent answer\cite{Fujikawa1}. 
Then, formula (\ref{formula}) is
nothing but
\begin{equation}
\ln {\mathcal J} \;=\; - 2 i \left.\alpha \, {\rm Tr}\,
\gamma_5 \right|_{reg}
\end{equation}
which then yields
to the known
result for the chiral anomaly~\cite{Fujikawa1},\cite{GMSS}
\begin{equation}
\langle \partial_\mu j^\mu_5 (x) \rangle = \left. 
\frac{\delta \ln {\mathcal J} }{\delta \alpha(x)} \right|_{\alpha = 0}= 
\left.2 i \, {\rm Tr}\,
\gamma_5 \right|_{reg}
 \end{equation}

We see that the choice of $\tanh({i{\mathcal D}_c}/{M})$ to construct  
 ${\mathcal D}$ through the defining equation
 (\ref{eq:defd}), which in turn allowed (written
as an infinite product) to the definition of the $D+1$
dimensional finite theory corresponds, at the level of regularized
quantities, to the choice of a  regulator which is close (but does not
coincide) with the usual adopted heat-kernel $\exp(-{\mathcal D}^2_c/{M^2})$
or zeta-function $\zeta({\mathcal D}_c,s)$ regulators.
 
\section{Conclusions}\label{sec:conc}
We have introduced a higher dimensional representation for regularized
Dirac and scalar field theories which preserves the symmetries of the
additional system, in spite of the fact that the regularization is
non-perturbative, namely, it yields results that can be shown to be
finite even without invoking power-counting arguments. The length of
the extra coordinate may be related to a fictitious `temperature',
although the resulting finite temperature QFT has of course unphysical
features, a property that indeed should be expected from any {\em
  regularized\/} theory.  This finite temperature representation has
the virtue that it automatically leads to the regulator masses one
should introduce when using an infinite number of Pauli-Villars
fields, as in~\cite{Frolov:ck}. Moreover,  it also leads to a natural 
extension of the regularization to scalar field theories.

Since the method leads naturally to a regularized {\em action}, it
allows for the study of the realization of symmetries and their
corresponding anomalies. In this respect, we have  discussed 
for the fermionic case, the
interplay between the   symmetries of the original theory
and the supersymmetry associated to the doubling of
fields in the regulated
theory. In particular, we have studied in details the case of chiral
anomalies, showing that the regularized action is invariant under
transformations which tend to the usual chiral ones when the regulator
is removed. At the quantum level, there is a nontrivial
anomalous Jacobian  which, as it was to be expected,   
is already regulated. The fact that this Jacobian 
is different from $1$ may be understood 
as a consequence of the breaking of the supersymmetry by the necessity of 
imposing different boundary conditions for the fields involved.

It should be remarked that the symmetries of the regularized theory
are non-local, a property that should be expected since it holds also
for the standard perturbatively regularized theories~\cite{sym1,sym2}.

\section*{Acknowledgements} 
C.~D.~Fosco is supported by CONICET (Argentina), and by a Fundaci{\'o}n
Antorchas grant. F.A.S is partially
supported
by UNLP, and ANPCYT(PICT 03-05-179),  Argentina.


\end{document}